\documentclass[twocolumn,showpacs]{revtex4}

\usepackage{epsfig}

\usepackage{dcolumn}
\usepackage{bm}

\def\optsection#1{}

\begin{document}

\title{Fine Structure of Dark Matter Halos and its Effect
on Terrestrial Detection Experiments}

\author{David Stiff and Lawrence M. Widrow}
 \affiliation{Department of Physics, Queen's University,
Kingston, K7L 3N6, Canada}
\altaffiliation{stiffd@astro.queensu.ca}
\altaffiliation{widrow@astro.queensu.ca}
\date{\today}


\begin{abstract}

Terrestrial dark matter detection experiments probe the velocity-space
distribution of dark matter particles in the vicinity of the Earth.
We present a novel method, to be used in conjunction with standard
cosmological simulations of hierarchical clustering, that allows one
to extract a truly local velocity-space distribution in exquisite
detail.  Preliminary results suggest a new picture for this
distribution which is decidedly non-Maxwellian but instead is
characterized by randomly positioned peaks in velocity space.  We
discuss the implications of these results for both WIMP and axion
detection experiments.

\end{abstract}

\maketitle

\optsection{Introduction} 

According to the cold dark matter (CDM) model of structure formation
galaxies possess extended massive halos of non-baryonic particles.
But while the CDM scenario can account for the rotation curves of
spiral galaxies and the morphology of structure on sub and
super-galactic scales it sheds little light on the fundamental nature
of dark matter.  For this, we must turn to dark matter detection
experiments.

WIMPs (weakly interacting massive particles) and axions are two dark
matter candidates for which direct detection in a terrestrial
apparatus is a distinct possibility.  Terrestrial detectors probe the
distribution function (DF) of dark matter particles in the lab.  The
standard WIMP detection experiment attempts to measure the energy
deposited when a WIMP scatters off a nucleus in the detector
\cite{goodman}.  By contrast the direct detection scheme for axions
relies on their coupling to photons: In a magnetic field, an axion can
convert into a photon whose energy will be equal to the total (rest
mass plus kinetic) energy of the axion \cite{sikivie}.

The phase space DF, $f\left ({\bf x},\,{\bf v}\right )$ provides a
complete description of a collisionless system such as a dark matter
halo.  Formally $f\left ({\bf x},\,{\bf v}\right )d^3x d^3v$ is the
number of particles in the six-dimensional phase space element
$d^3xd^3v$ centered on the phase space point $\left ({\bf x},\,{\bf
v}\right )$.  A detector of volume ${\cal V}$ at position ${\bf x}_0$
probes the velocity space DF $g_{{\bf x}_0}\left ({\bf v}\right
)\equiv {\cal V}^{-1}\int_{\cal V} d^3x f\left ({\bf x},\,{\bf
v}\right ) $.  Since detectors are very small, $g_{{\bf x}_0}\left
({\bf v}\right )=f\left ({\bf x_0},\,{\bf v}\right )$ to extremely
high accuracy.

The standard assumption for direct detection experiments is that $g$
can be approximated by a Maxwellian with a cut-off at the escape speed
of the Galaxy.  Our aim is to explore the extent to which this
assumption breaks down within the CDM scenario for structure
formation.  In the CDM model, dark halos are assembled through
hierarchical clustering wherein small-scale objects collapse first and
merge to yield systems of increasing size.  Originally, it was
believed that the substructure one might have expected from
hierarchical clustering was erased by physical processes such as
violent relaxation and tidal stripping \cite{white}.  However, recent
high resolution simulations indicate that dark halos are in fact
clumpy with a significant fraction of the mass bound in undigested
lumps and tidal debris \cite{klypin, moore}.

The question is whether substructure in the density field of the halo
translates into structure in the local velocity-space DF $g$.  If so,
there may be important implications for terrestrial detection
experiments.  (The implications of triaxiality, rotation, and a radial
profile that differs from that of an isothermal sphere have been
considered by various authors \cite{kamion2, ullio, green, copi}.)  In
the CDM scenario, the dark matter in the vicinity of the Earth is a
superposition of bound clumps and tidal streams that were accreted by
the Galaxy over its entire formation history.  From the standpoint of
direct detection experiments, particles from recent accretion events
are the most interesting since they will have a bulk galactocentric
velocity at the position of the Earth of order the escape speed of the
Galaxy ($\sim 500\,{\rm km s^{-1}}$) or roughly twice the mean speed
of the widely used Maxwellian model.  Moreover, even if most of the
particles in these clumps have been stripped by the tidal field of the
Galaxy, the tidal debris will have little time to phase mix and
therefore be relatively coherent.  Coherent streams of high velocity
particles may produce significant features in detection spectra even
if they constitute a small fraction of the total local dark matter
density \cite{stiff,freese}.  (See, however, \cite{helmi} where the
authors reach a different conclusion.)

The Maxwellian model may also be challenged by the following line of
reasoning.  The intrinsic velocity dispersion of CDM is negligible and
the distribution function $f\left ({\bf x},\,{\bf v}\right )$ is most
appropriately thought of as a three-dimensional sheet in
six-dimensional phase space.  Prior to the epoch of structure
formation, this sheet is nearly perfectly flat: the velocities of
particles at a particular point in space are given by the Hubble flow
plus small perturbations.  As structure formation proceeds, $f$
retains its three-dimensional character but is ``curled up'' through
the process of gravitational collapse.  $g\left ({\bf v}\right )$ is
the value of $f$ on a three-dimensional surface that defines the
detector, i.e., a surface that spans the three velocity-space
directions at the position of the detector.  Since the intersection of
two three-dimensional surfaces (one flat and the other highly
contorted) is a collection of points, $g$ will be a collection of
discrete peaks in velocity space.  These peaks will be randomly
distributed though will likely exhibit clustering on different
(velocity-space) scales.  (For a general discussion of the
dimensionality of phase-space structures as it relates to astronomical
observations, see Ref.\cite{tremaine}).

Fine structure of this type has been studied within the context of
self-similar spherical infall models \cite{sikivie2}.  These models
assume that the initial density perturbation and angular momentum
distribution are scale free and spherically symmetric.  Given these
assumptions, one can solve semi-analytically for the entire formation
history of a halo and predict the positions of the velocity-space
peaks which, in turn, may lead to a distinctive pattern of features in
energy spectra obtained by axion and WIMP detectors \cite{sikivie2,
copi}.

Velocity-space structures in hierarchically formed halos can be
studied using numerical simulations as follows: Consider a Milky
Way-size simulated halo with $N$ particles and select a location that
has the characteristics of the solar system (for example, a point
$8.5\,{\rm kpc}$ from the center of mass of the halo).  Choose a
volume ${\cal V}$ centered on this point.  The $N'$ particles inside
this volume provide a monte carlo realization of $g({\bf v})$.  This
procedure has been employed by various groups with some success
\cite{moore2,helmi}.  However, even the largest simulations are unable
to resolve velocity space structures deep inside highly evolved halos.
The limit on the velocity space resolution due to the finite number of
particles in ${\cal V}$ is $\Delta v/\sigma \sim N'^{-1/3}$.  Velocity
space resolution is also limited by the finite size of ${\cal V}$.
Let $L$ and $\Phi$ be respectively the size of the halo and depth of
its gravitational potential, $L'$ be the size of ${\cal V}$, and
$\Delta\Phi$ be the variation in $\Phi$ across ${\cal V}$.  The limit
to the resolution in velocity space due to the finite size of ${\cal
V}$ is $\Delta v/\sigma\sim \Delta\Phi/\Phi\sim L'/L \sim \left (
N'/N\right )^{1/3}$.  We minimize $\Delta v/\sigma$ by balancing the
two effects (large ${\cal V}$ for particle number, small ${\cal V}$
for locality) to find the optimal choice $N'\simeq N^{1/2}$.  The
implication is that for a simulation of $N$ particles, only $O\left
(N^{1/2}\right )$ are available to map the local velocity space
distribution function.

This paper introduces a new technique, to be used in conjunction with
standard cosmological simulations, which allows one to map $g({\bf
v})$ at a single point within a dark halo.  The method relies on test
particles which, by design, reach the desired point (the position of
the would-be detector) in the final timestep of the simulation.
Through an iterative process, these test particles allow one to locate
the points where the phase space sheet describing the dark matter
distribution intersects the phase space sheet describing the detector.
Our method has a resolution in velocity space of $\Delta v/\sigma \sim
N^{-1/3}$ as compared with $\Delta v/\sigma \sim N^{-1/6}$ for the
standard approach.  Thus, with $N=10^6$, we can achieve a resolution
that would have previously required $10^{12}$ particles.

\optsection{Reverse-Run Method} 

The starting point for our algorithm is a simulation of the formation
of a dark matter halo from standard CDM initial conditions.  In the
final frame of the simulation, we identify an appropriate location for
a detector and lay down a uniform grid (i.e., three-dimensional
phase-space sheet) of massless test particles which spans
velocity-space at that location.  We next evolve both the test
particles and original simulation particles backward in time to the
initial frame of the simulation.  The backward evolution is
accomplished by simply reversing the timestep.  During this ``reverse
run'' the phase-space sheet defined by the test particles will fold
and curl.  At the initial time, we locate the intersection points of
the test-particle phase-space sheet and the phase-space sheet that
defines the initial dark matter DF.  These intersection points map
directly into points in velocity space at the position of the detector
(ie. the initial velocity-space distribution of the test particles).
The value of the DF is calculated at each crossing point, and by
Liouville's theorem, the DF of the test particles at the location of
the detector can be determined.

One may think of the test particles as providing the means to
interpolate the DF in the region of the detector.  For the
interpolation to make sense, the system must reverse properly, i.e.,
the positions of the simulation particles at the end of the backward
integration should be within some small distance of their actual
initial positions.  In essence, reversibility is equivalent to the
condition that the dark matter DF, as described by the simulation
particles, maintains its character as a three-dimensional structure in
phase space.  Chaos, which is almost certainly present in the orbits
of the simulation particles, combined with round-off errors, has the
potential to spoil the reversibility of the time integration and
destroy the integrity of the phase-space sheet.  To apply our
technique to a simulation, it must be verified that it does reverse
properly. We have found that the simplest means of suppressing chaotic
behavior and achieving reversibility is to increase the softening
length above what is normally used in cosmological simulations.

At this point, the reader may suspect a swindle.  If our method
requires a larger softening length than is normally used, how can we
claim a substantial improvement over the standard N-body approach.  Of
course, chaos also affects the results of the standard method.  We
suspect that the approximately Maxwellian DFs found in previous
studies \cite{helmi, moore2} arise in part from the use of a finite
volume to measure the velocity-space DF at a single point together
with the fact that the dark matter phase-space sheet is not adequately
modelled.  Our interpolation scheme overcomes the finite volume
problem and makes explicit the requirement that the phase space sheet
be properly modelled.

Typically we begin with the same number of test particles as
simulation particles so that the additional computation time is
increased by a factor of less than 2.  Higher resolution is achieved
by an iterative procedure -- regions surrounding intersection points
are resampled by a finer grid of test particles and the reverse-run
algorithm is repeated.  Our simulations use an implementation of
Dehnen's $O(N)$ algorithm \cite{dehnen} modified to efficiently
incorporate test particles. Our technique however, is independent of
the code used as long as it is time-reversible and allows test
particles. The softening kernel is chosen such that
the gravitational force goes to zero at zero separation but joins
smoothly with Newtonian gravity at the softening length.  In this
manner, a test particle and a real massive particle with exactly the
same phase-space coordinates will follow the same trajectory.

\optsection{Spherical Cold Collapse} 

As a demonstration of the power of our algorithm, we first examine the
collapse of a spherically symmetric density perturbation reminiscent
of the self-similar collapse considered in \cite{sikivie2} and
references therein. The perturbation is a monotonically decreasing
function of radius so that the inner parts of the system collapse
first. The simulation is run from $t=0$ to $t=12$ (in units with
$M_{\rm total}=100$ and $G=1$) with
approximately $33,000$ particles and uses the same N-body code that
will be used in subsequent, more realistic simulations. As the
simulation is done in physical coordinates, all particles have an
initial Hubble velocity and comoving softening length of 1.7 
(Plummer equivalent) was used.

\begin{figure}
\centerline{\epsfig{file=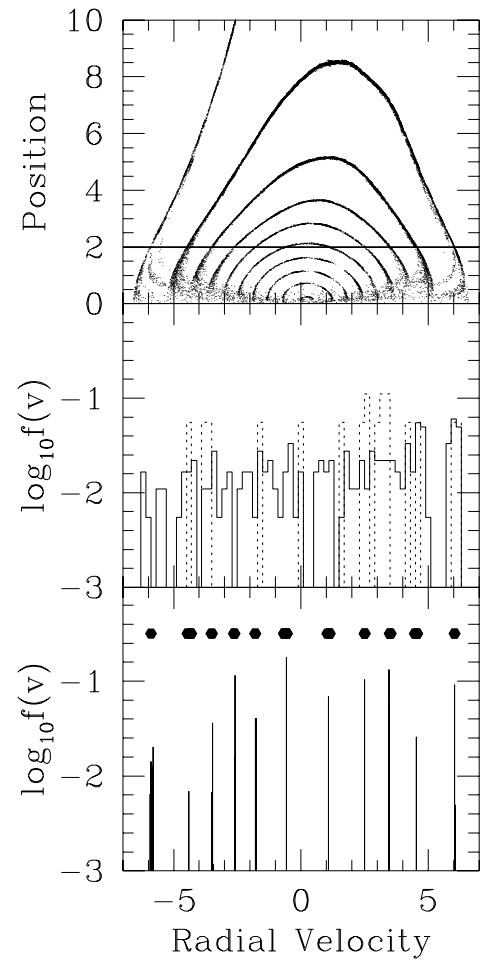,width=2.5in}}
\caption{Top: The $r-v_r$ diagram at the final time.  Middle: The
  velocity distribution 
  determined from the 
  forward run with a large (solid) and small (dashed) volume at
  $x_0$. Bottom: The 
  reconstructed velocity distribution from our reverse-run
  technique. The dots along the top of the figure
  indicate the location of streams from the
  forward run when spherical symmetry is applied.}
\label{fig:cc2}
\end{figure}

The top panel of Figure \ref{fig:cc2} shows the final DF (radius $r$
vs.\,radial velocity $v_r$).  The fine structure (thin streams of
particles) should be viewed as phase space sheets where we have
projected out the $\theta$ and $\phi$ directions.  In the absence of
spherical symmetry, the three-dimensional nature of the DF would still
be present but not readily apparent in an $r-v_r$ plot.  Indeed,
toward the center of the halo, where two-body scattering has destroyed
spherical symmetry, the structures appear to dissolve.

A detector at $x_0 = (2,0,0)$ (horizontal line in the top panel of
Figure \ref{fig:cc2}) should measure $11$ distinct velocity-space
peaks.  The middle panel of Figure \ref{fig:cc2} shows the results for
the velocity space distribution function obtained using the standard
N-body method.  Two different volumes, one of
radius $\Delta r = 0.89$ 
and containing 182 particles (approximately
$N^{1/2}$) and the other of radius $\Delta r = 45$ 
and containing 18 particles were used to generate these results.  With
a large volume, one obtains a broad distribution and sees little
evidence for discrete structures.  A smaller volume picks out only a
few of the peaks, and with very poor resolution.

The bottom panel shows the results of the reverse-run method.  All 11
peaks are detected.  The plot is an actual histogram: the thickness of
the peaks reflects the actual velocity-space resolution obtained by
our method.  Along the top of the plot are the positions of the peaks
as measured using the N-body approach but exploiting spherical
symmetry, i.e., taking all particles close to $r=2$.  We see that the
peaks from the reverse run method are precisely where they should be.

\optsection{Milky Way Simulation} 

\begin{figure}
\centerline{\epsfig{file=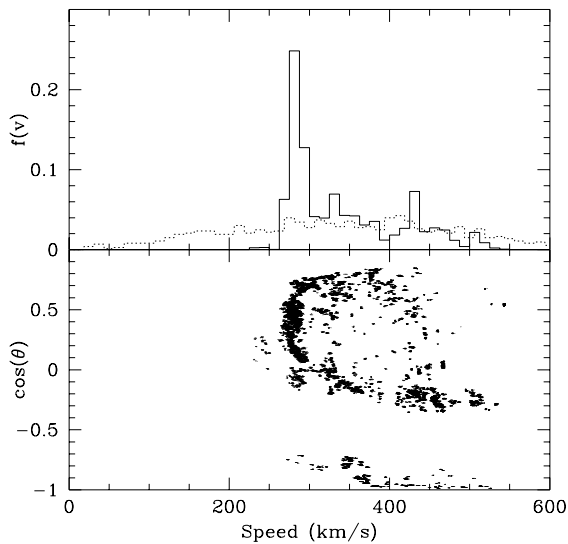,width=3in}}
\caption{Top panel: Speed distribution from the forward run with $N^{1/2}$
  particles (dotted) and from our reverse-run technique (solid). Lower
  panel: Distribution of angle between the local bulk motion and the
  crossing points.}
\label{fig:kgvel}
\end{figure}

The formation of a dark halo in the hierarchical clustering scenario
is modelled by augmenting the spherically symmetric simulation
considered above with a power-law spectrum ($n=-2$) of density and
velocity perturbations (see, for example, reference \cite{katz}). The
particle number was increased to $5 \times 10^5$ and the softening set
to 20 kpc Plummer equivalent.  The simulation started at $z=44$ and
evolved to the present in 1000 timesteps. The total system mass was
$10^{13} M_\odot$. It should be noted that this simulation is only
designed to qualitatively model the formation of the Milky Way so
rigorous normalization of the power spectrum was not performed.
Nonetheless, the resulting system consists of a massive central galaxy
($M \sim 10^{12} M_\odot$) which partly formed through the merger of
smaller objects and several dwarf satellites galaxies surrounding it.
One can view the particle distribution and evolution of the simulation
elsewhere \cite{stiffwebsite}.  Figure \ref{fig:kgvel} shows the speed
distribution and a scatter plot in the two-dimensional phase space of
speed and angle where the angle is measured with respect to some
arbitrary direction in the halo.  The latter may be of interest to
detectors with directional sensitivity such as the DRIFT detector
\cite{drift}.

\begin{figure}
\centerline{\epsfig{file=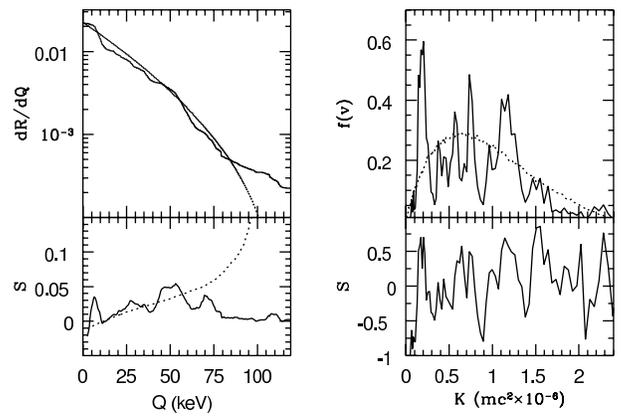,width=3.4in}}
\caption{Left Top: WIMP recoil energy spectra from the reverse-run
(solid line) and a numerical Maxwellian distribution designed to match
the average properties of the reverse-run results (dotted line). Left
Bottom: Seasonal modulation factor $S(E)$. Right Top: Axion energy
spectra from the reverse-run (solid line) and the same Maxwellian
(dotted line).  Right Bottom: Seasonal modulation factor $S(E)$.}
\label{fig:wimpspectra}
\end{figure}

The top-left panel of Figure \ref{fig:wimpspectra} gives an example of a
recoil spectra that would be obtained by our fictitious detector.
Since terrestrial detection experiments operate in the rest frame of
the Earth we first shift the velocity distribution by $220\,{\rm
km\,s^{-1}}$.  (We have not attempted to model the Galactic disk and
therefore the direction of this shift is arbitrary.)  In addition, we
include the small velocity-space shift due to the Earth's motion
around the Sun.  It is this shift that gives rise to the famous
seasonal modulation effect considered to be so important to
terrestrial detectors \cite{freese2}. The seasonal modulation factor,
$S(E) \equiv $ [Rate(June) - Rate(December)]/[Rate(June) +
Rate(December)], (see Ref.\,\cite{stiff}) is shown in the lower-left
panel of the Figure.

The recoil spectrum is characterized by a sequence of steps as one
goes up in energy.  As noted in Refs. \cite{copi,stiff,freese} peaks
in the speed distribution translate to step functions in a recoil
spectrum.

Our ability to identify velocity space structure may be considerably
better if dark matter is composed of axions.  The standard axion
detector measures the energy of the axion (kinetic plus rest mass) and
therefore peaks in velocity space translate into peaks in the energy
spectrum.  In addition, since these experiments rely on the
orientation of the magnetic field, one will have directional
sensitivity and therefore be able to map out the full $g({\bf v})$.
Figure \ref{fig:wimpspectra} provides an example of an axion spectrum
obtained by again shifting the velocity distribution to a terrestrial
frame.

\optsection{Conclusions} 

The reverse-run algorithm provides the means to probe the local
velocity-space structure of a simulated halo with a resolution far
greater than is permitted through traditional methods.  In short, our
method allows us to obtain the velocity-space distribution at a given
point without having to average over the surrounding region.

Our results must be viewed as preliminary since our simulations focus
a small region of the Universe and therefore do not take into account
cosmological tidal fields.  Moreover, no attempt is made to model disk
formation.  However, our method is completely general and may be
applied to more realistic simulations of the Local Group or Local
Supercluster.

The preliminary results do suggest a new picture for the local
velocity-space distribution function, one which is characterized by
discrete peaks.  This result is potentially of great importance to
terrestrial detection experiments which rely on models of this
distribution function to develop search strategies and interpret
experimental results. For WIMP searches, the deviations from
Maxwellian and clumpy distributions are likely not significant for the
current generation of detectors. However, current axion detectors have
exceptional energy resolution on the order of $\Delta E/(m_a c^2) \sim
10^{-11}$ \cite{asztalos}. This resolution is sufficient to be
sensitive to seasonal and even diurnal variations in the spectrum
which can be used to fully resolve the velocity-space distribution of
axions. Future generations of WIMP detectors with directional
sensitivity, or accurate seasonal spectra may also yield detailed
features of the dark matter distribution.

  We are in fact led to the tantalizing
conclusion that if terrestrial experiments are able to detect dark
matter, then the results will provide a wealth of information on the
structure and formation history of the dark halo.



We thank R. Henriksen and T. Merrall for useful conversations and
J. Frieman and T. Noble for comments on an early version of the
manuscript. This work was supported, in part, by the National Science
and Engineering Research Council of Canada.

\end{document}